\documentclass{emulateapj}
\shorttitle{Implications of Spin-Orbit Misalignment}
\shortauthors{Thorsett, Dewey, \& Stairs}
\begin{document}

\submitted{Submitted to ApJ Aug. 24, 2004}

\title{Studies of the Relativistic Binary Pulsar PSR B1534+12. II. Origin and
Evolution}

\author{S. E. Thorsett, R. J. Dewey}
\affil{Department of Astronomy and Astrophysics, University of
		California, Santa Cruz, CA 95064}
\email{thorsett@ucolick.org}
\email{dewey@astro.ucsc.edu}
\and
\author{I. H. Stairs}
\affil{Department of Physics and Astronomy, 
		University of British Columbia,
		Vancouver, BC V6T 1Z1, Canada} 
\email{istairs@astro.ubc.ca}

\begin{abstract}
We have recently measured the angle between the spin and orbital angular momenta
of PSR B1534+12 to be either $25\pm4^\circ$ or $155\pm4^\circ$. This
misalignment was almost certainly caused by an asymmetry in the supernova
explosion that formed its companion neutron star. Here we combine the
misalignment measurement with measurements of the pulsar and companion masses,
the orbital elements, proper motion, and interstellar scintillation. 
We show that the orbit of the
binary in the Galaxy is inconsistent with a velocity kick large enough to
produce a nearly antialigned spin axis, so the true misalignment must be
$\sim25^\circ$. Similar arguments lead to bounds on the mass of the companion
star immediately before its supernova: $3\pm1M_\odot$. The result is a coherent
scenario for the formation of the observed binary. After the first supernova
explosion, the neutron star that would eventually become the observed pulsar was
in a Be/X-ray type binary system with a companion of at least 10--12\,$M_\odot$.
During hydrogen (or possibly helium) shell burning, mass transfer occurred in a
common envelope phase, leaving the neutron star in a roughly half-day orbit with
a helium star with mass above $\sim3.3M_\odot$. A second phase of mass transfer
was then initiated by Roche lobe overflow during shell helium burning, further
reducing both the helium star mass and orbital period before the second
supernova. Scenarios that avoid Roche lobe overflow by the helium star require
larger helium star masses and predict space velocities inconsistent with our
measurements. The companion neutron star experienced a velocity kick of
$230\pm60$~km/s at birth, leading to a systemic kick to the binary of
$180\pm60$~km/s. The direction of the kick was roughly opposed to the
instantaneous orbital velocity of the companion, but the kick angle is largely
unconstrained.
\end{abstract}


\section{Introduction}

The handful of known short-period double neutron star binaries are of interest
to physicists and astrophysicists working in a wide variety of fields. Most
famously, studies of two of these systems have been used as laboratories to test
aspects of fundamental gravitation theory that are inaccessible in the solar
system, including the prediction by \citet{ein16} of the production of
gravitational radiation \citep{tw89,sttw02}. Such binaries, in their late stages
of gravitational inspiral, are among the most likely sources to be detected by
LIGO and other gravitational wave observatories \citep[and references
therein]{aaa+04}. They yield high-precision mass measurements that have been
used to study the neutron star equation of state \citep{tc99}. Finally, they are
important long-lived relics of binary evolution that preserve in their orbital
elements a memory of short-lived mass transfer phases that are rarely if ever
observed in progress \citep[e.g.,][]{bv91}.

A complicating factor in understanding the birthrate of these binaries has been
constraining asymmetries in supernova explosions. Isolated radio pulsars are
observed to have high space velocities, interpreted as evidence that newborn
neutron stars receive a momentum kick at birth \citep{dc87,ll94}. There is some
evidence that kicks given to stars in binaries are smaller
\citep{ctk94,cc97,py98,hb99}, but it is unclear whether this is a selection bias
related to the probability of binary survival, or whether these stars truly have
a different kick distribution, perhaps because of a different angular momentum
at collapse. Since kicks introduce new degrees of freedom in the change of the
binary elements after the supernova event, understanding kicks is also important
for estimating the pre-supernova orbital properties. Finally, asymmetries can
produce misalignments between the spin and orbital angular momenta in the
binary, which can lead to changes in the gravitational-wave signature at merger
\citep{acst94,gkv03}. Misalignment also makes possible the study of general
relativistic (geodetic) precession, providing a qualitatively new test of strong
field gravity theories \citep{dr74,bo75,bo75b,eh75,dt92,sta04}.

Recently, we have used precession measurements (together with the assumption
that general relativity correctly describes the kinematics of neutron star
binaries) to measure the misalignment angle in the 
PSR B1534+12 binary \citep{sta04}. By
combining this result with our previous high precision measurements of the
orbital elements, component masses, system distance, and proper motion, and with
a measurement of the orientation of the orbit from interstellar scintillation
studies, we are
able to tightly constrain the magnitude and direction of the asymmetric velocity
kick imparted to the pulsar's companion when it was formed. Three unknown
parameters---the line-of-sight velocity of the binary, the pre-supernova orbital
size, and the pre-supernova mass of the companion---are all tightly constrained.
These in turn strongly constrain the evolutionary history of the binary, at
least during the period between the first and second supernovae, as well as
the supernova asymmetry.

Our work here draws on previous studies of several relativistic binaries,
particularly of this system and of PSR B1913+16. However, the quality and extent
of the data for PSR B1534+12 now far exceeds what is available for the other
binaries, allowing much more robust conclusions to be drawn. We describe these
data in \S\ref{sec:obs}. In \S\ref{sec:theory} we review the generic 
evolution of a compact
binary system before, during, and after the second supernova 
event. In
\S\ref{sec:results}, we combine observations with theory to constrain the
particular history of the PSR B1534+12 binary.

\section{Observations of PSR B1534+12}\label{sec:obs}

\begin{figure}[t]
\plotone{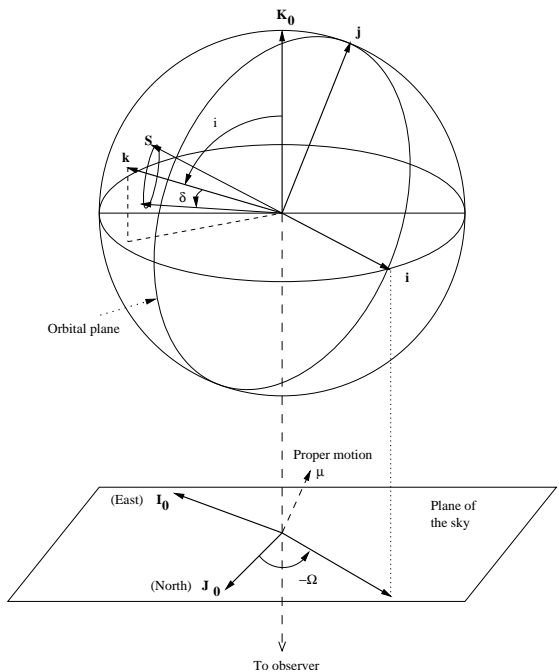}
\caption{\label{fig:geom}The geometry of the PSR B1534+12 orbit. The orbital
angular momentum vector is $\mathbf{k}$, and the spin axis is $\mathbf{S}$. The
precession cone, with half-opening angle $\delta$, is shown. The vector
$\mathbf{i}$ is directed towards the ascending node.}
\end{figure}

As we will see below, the current orbit and kinematics of the PSR B1534+12
binary were largely determined by the properties of the binary just prior to the
second supernova event (which created the neutron star companion to the observed
pulsar) and by the mass loss and momentum imparted to the system during the
second supernova. Neither the size nor the eccentricity of the orbit have changed
significantly since that event, while the binary has moved ballistically through the
Galactic gravitational potential and gyroscopically maintained the direction of
its orbital angular momentum vector. A wealth of data fully constrain the
current orbital properties, including the orientation of the orbit on the sky.
(See Fig.~\ref{fig:geom} for a reference to angles describing the
orientation.)
The distance and transverse components of the binary velocity are also well
constrained. Only the radial velocity and the age cannot be directly determined,
though a rough upper limit can be placed on the time since the second supernova.

\subsection{Orbital elements and masses}

Radio pulsar timing has produced exquisitely accurate estimates of the orbital
elements of the PSR B1534+12 binary. The system is essentially a single-line
spectroscopic binary, with the Doppler shift of the pulsar period replacing the
usual Doppler shift of a stellar spectral line. Although only line-of-sight velocity
variations can be observed through timing, the measurement of 
general-relativistic corrections to the Keplerian orbital equations allow the
orbital inclination and component masses to be determined. 
The timing experiments, analysis, and
results have been described in extensive detail in \citet{sttw02}. For our current
purposes, it is sufficient to know the orbital period 
$P_b=0.421$~days, eccentricity $e=0.274$, relative semimajor axis $a=7.62$~light
seconds, orbital inclination $\sin i=0.975$, pulsar mass $m_1=1.333M_\odot$, and
companion mass $m_2=1.345M_\odot$. In all cases, the uncertainty is small
relative to the last quoted digit.

\subsection{Misalignment of spin and orbital angular momenta}

Misalignment between the pulsar spin axis and the orbital angular momentum axis
causes the spin axis to precess around the orbital axis at a rate of
$0.51^\circ$/yr. This so-called geodetic precession is a simple relativistic
consequence of the parallel transport of the spin vector in curved space; it is
a consequence of the same angle deficit that produces precession of the
periastron of the orbit.

Precession of the pulsar spin axis causes the impact angle between the
observer's line of sight and the pulsar's magnetic pole to change, with an
amplitude that depends on the misalignment angle $\delta$ between the spin and
orbital angular momenta. This shifting viewing angle causes changes in the
apparent pulsar beam shape and polarization properties.

Recently, we have used a long-term study of the polarization and beam shape to
measure $\delta=25.0\pm3.8^\circ$ \citep{sta04}.  
There is a discrete ambiguity that corresponds
to reversing the spin direction, so $\delta=155.0\pm3.8^\circ$ is also allowed.

\subsection{Position and velocity}

In addition to the orbital elements, timing measurements provide very accurate
estimates of the pulsar position and proper motion, through the annual
modulation of the observed pulse period caused by the Earth's motion. In the
case of PSR B1534+12, the distance can also be estimated from a measurement of
the apparent orbital period derivative (after correction for the contribution
due to gravitational radiation emission)---which is caused by a combination of
acceleration in the Galactic potential and apparent acceleration along the line
of sight that arises from the transverse motion.

The binary's current position is well out of the Galactic plane, at
$l=19.8^\circ$, $b=48.3^\circ$. The distance is $d=1.02\pm0.05$~kpc. The
observed proper motion is $\mu_{\rm RA}=1.3$~mas/yr, $\mu_{\rm dec}=-25.1$~mas/yr
\citep{sttw02}. The total proper motion is $\mu=25.2$~mas/yr, directed towards a
Galactic position angle $239.2^\circ$, which at the known distance implies a
transverse velocity (uncorrected for Solar motion) 
of 122~km/s. In contrast to the usual case with a spectroscopic
binary, there are no direct constraints available on
the radial component of the space velocity as the rest-frame spin period of the pulsar
is unknown.

In projection, the pulsar appears to be moving towards the plane, with a $b$-component velocity
$-62$~km/s. However, a radial velocity component of just 56~km/s is sufficient
to reverse the apparent motion, so it is unknown whether the pulsar is presently
moving towards or away from the plane.

\subsection{Scintillation}

Timing studies, which measure the component of velocity along the line sight,
are insensitive to rotations of the binary orbit around the line of sight, and
hence the angle $\Omega$ of the ascending node on the plane of the 
sky.\footnote{Note
that we use the standard convention in binary pulsar studies and define
the ``ascending node'' as the node at which the observed pulsar is moving
away from the observer.} This 
angle is accessible through proper-motion-induced changes in the projected
orbital size, with $\dot x/x=\mu\sin j\cot i$, with $i$ the orbital inclination
and $j$ the angle between the
proper motion direction and $\Omega$ \citep{ajrt96}. Because $\mu$ is small for
PSR B1534+12, this has not yet been measurable.

In principle, some constraint on the direction of the pulsar's spin axis could be
obtained from polarization studies, if corrections are made for Faraday rotation
and observations are properly calibrated against an absolute reference source.
This has also not yet been done for PSR B1534+12.

An alternative approach that has been successful for this source comes from
study of diffractive scintillation. The propagation of the pulsar signal through
the inhomogeneous interstellar medium produces a pattern (``screen'') of
intensity variations caused by constructive or destructive interference. The
speed at which this screen sweeps past the observer is proportional to the
vector sum of the Earth's motion, the binary's proper motion, and the pulsar's
orbital motion, and is inversely proportional (at least statistically) to the
decorrelation timescale for intensity variations \citep{lyn84,dcww88}.

By measuring the decorrelation timescale as a function of orbital phase,
\citet{bplw02} were able to measure $\Omega=70\pm20^\circ$ or $290\pm20^\circ$, 
reckoned north through east. The solutions correspond to $\cos i<0$ and $\cos i>0$,
respectively;\footnote{Although \citet{bplw02} do not explicitly identify which
solution corresponds to which sign of $\cos i$, our independent analysis has been
confirmed by Bogdanov, private communication.} 
only the second is consistent with our polarization studies \citep{sta04}.
Although the
uncertainties remain large, we will see that the resulting constraint on the direction of
the orbital angular momentum is useful.

\subsection{Age}

A simple estimate for the age of the binary since the second supernova is given
by the characteristic age of the observed pulsar: $\tau_c\equiv P/2\dot
P=0.25$~Gyr. The true age $\tau$ is unknown. If the pulsar was born at or below
the spin-up line and has evolved by magnetic dipole braking (with braking index
$n=3$), then $\tau\stackrel{<}{\sim} 0.21$~Gyr \citep{acw99}. The age could be significantly
less than this if the pulsar spin-up ended before the spin-up line was reached,
or if mass transfer ended (and spin-down began) well before the second
supernova, and could be somewhat larger if $n<3$. As will be discussed in
\S\ref{sec:kinematics}, ages as small as 10~Myr allowed if the binary is just
leaving the plane for the first time. If the system is truly that young, the
companion star may still be an active radio pulsar, which cannot be ruled out if
the Earth lies outside the path of its lighthouse beam. We believe it is
unlikely that the age of the binary is much larger than the characteristic age.
The direct constraints on effective braking indices for pulsars with
intermediate magnetic fields, like PSR B1534+12, is limited. However,
\citet{bfg+03} have found that characteristic ages and kinematic ages agree well
in a set of high field pulsars, and in a sample of millisecond pulsars (with
smaller magnetic fields), \citet{ctk94} found that half had characteristic ages
larger than a Hubble time, implying that characteristic ages were typically an
overestimate of true age. Without further information, we take
$2.1\times10^8$~yrs as a rough upper limit to the age of the system, but none of
our conclusions depend strongly on that limit.

\section{The evolution of the PSR B1534+12
binary}\label{sec:theory}

In general terms, the formation of close double neutron star binaries like PSR
B1534+12 is well understood \citep[see][for example, for a review]{bv91}. It
begins with a first supernova in a binary star system, in which the neutron star
that is now observed as a pulsar was born. After a possible Be/X-ray binary
phase, a period of mass transfer occurs: ``case~B'' if it occurs by Roche lobe
overflow during hydrogen shell burning, as the companion first climbs the giant
branch, or ``case~C'' if it occurs during helium shell burning. (``Case~A''
transfer, while the companion is still on the hydrogen main sequence, is thought
to lead to destruction of the binary through merger.) During the ensuing common
envelope evolution, the system spirals together, leaving a binary containing a
neutron star and a bare helium star. After case~B transfer, the helium star is
unevolved. After case C transfer, the helium star has already developed a
carbon-oxygen core. A second stage of mass transfer may occur, which is again
called case~A, B, or C if it occurs during helium core burning, helium shell
burning, or carbon burning, respectively. The overall mass-transfer history is,
for example, case BB if a first phase occurred during hydrogen shell burning and
a second during helium shell burning. In any case, during mass transfer the
orbit is circularized and the pulsar is spun-up, or ``recycled,'' and the pulsar
spin axis  aligned with the orbital angular momentum. Finally, a second
supernova explosion produces the companion neutron star, which may live for
$\sim10^7$~yrs as a pulsar. (The faster, recycled pulsar can live $\stackrel{>}{\sim}
10^9$~yrs, so most observed double-neutron star binaries are similar to B1534+12, with
only the recycled pulsar visible, and only in rare cases like the recently
discovered PSR J0737$-$3039 \citep{lbk+04} can both pulsars be seen.)

\citet{bro95} proposed an alternative scenario in which the initial inspiral
of two stars of similar mass produces a double helium star binary. Though
different in the early phases of evolution, this scenario still requires mass
transfer after the first supernova in order to recycle the observed pulsar, so
the system just prior to the second supernova is again expected to be a close,
circular neutron-star--helium-star binary with the pulsar spin and orbital
angular momenta aligned. We will not further discuss
this possible evolutionary track, except to note that because the late evolution
is very similar to the standard scenario most of our conclusions would be
unchanged.

At this point, it is important to emphasize that the immediate progenitors of
systems like PSR B1534+12---that is, close binaries that contain a neutron star
and bare helium star---are not observed, with the possible exception of Cyg X-3,
which appears to be a Wolf-Rayet star in orbit around either a neutron star or a
black hole \citep{vcg+92,fhp99}. Our primary observational constraints on the
late stages of the standard evolutionary scenario are the properties of the
resulting double neutron star binaries, including their masses, orbital
elements, and space motions. In the case of PSR B1534+12, a number of studies
have already been done \citep{ysn93,fk97,acw99,fwb02,dp03b,wkh04}. To these, we now add
improved estimates of the pulsar distance and proper motion \citep{sttw02}, 
measurement of the spin-orbit misalignment \citep{sta04}, and measurement of
the line of nodes angle \citep{bplw02}, which together
strongly constrain the magnitude of any asymmetric kick applied to the companion
star during the supernova in which it was born.

\subsection{The pre-supernova binary}

A number of constraints on the properties of the pre-supernova binary come from
the size of the helium star companion, the core mass required to form a neutron
star, and the possible stellar evolution and mass transfer histories that could
produce the close binary. A detailed study of the evolution of close helium star
binaries has recently been carried out by \citet{dpsv02} and \citet{dp03b}, and
we draw heavily from their work.

The small size of the pre-supernova binary---far too small to contain two main
sequence stars---immediately implies that the orbit went through at least one 
mass transfer and inspiral phase. In the standard model, the system after the
first supernova is a Be/X-ray type binary that spirals together in a common
envelope phase during either case B or case C mass transfer. (As noted above,
an alternate model posits inspiral during a double helium star phase before the
first supernova.)

The subsequent evolution depends on the helium star mass and whether a second
mass transfer phase occurs. As discussed in \citet{dpsv02}, case BA mass
transfer, during helium core burning, never results in a double neutron star
binary. Low mass stars lose enough mass that they leave CO white dwarf remnants,
rather than collapsing to neutron stars. High mass stars experience dynamically
unstable Roche lobe overflow, and probably coalesce with the neutron star to
form an isolated black hole.

Case BB mass transfer, during helium shell burning, results in the removal of
the helium star's envelope and the formation of a CO or ONe white dwarf if the
zero age helium star mass is less than $2.5M_\odot$. Stars with mass less than
$3.3M_\odot$ (or $3.8M_\odot$ in orbits below about six hour period) spiral in
during a common envelope phase, and probably produce very compact (15 minute)
double neutron star binaries, which have very short merger times and are
therefore difficult to detect \citep{dp03b}. More massive stars, up to about
$6.5M_\odot$, avoid a common envelope during Roche lobe overflow, and hence
avoid catastrophic inspiral. After mass transfer, these systems have orbital
periods of hours to days, and pre-supernova core masses between about 2.2 and
$4M_\odot$.  As we will see, these are progenitors for systems like PSR
B1534+12. 

Larger helium stars don't swell significantly as they evolve, so avoid Roche
lobe overflow. Binaries containing such massive He stars never experience Roche
lobe overflow, and are also possible progenitors of double neutron star
binaries. In this case, the spin-up of the observed pulsar occurs by accretion
from the helium star wind. In this scenario, a lower limit on the zero age
helium star mass comes from the requirement that it never expand into contact
with its Roche lobe, and an upper limit comes from the requirement that its zero
age radius fit within the Roche lobe. In the specific case of PSR B1534+12,
\citet{dp03b} have found that although a fairly wide range of zero age helium
masses are possible (up to about $13M_\odot$), after wind losses there is only a
very limited range of pre-supernova mass possible: $5\pm0.5M_\odot$. As we will
see, supernovae in these systems can reproduce the observed orbit of PSR
B1534+12, but only with a kick too large to be consistent with the observed
space velocity. There are no possible scenarios for the formation of PSR
B1534+12 that begin with case C transfer and then avoid Roche lobe overflow.

\subsection{The second supernova event}

We denote the orbital eccentricity, relative semimajor axis, and total mass
before the second supernova by $e_i$, $a_i$, and $M_i=m_{1}+m_{2i}$, where $m_i$
is the mass of the observed pulsar and $m_{2i}$ is the presupernova mass of the
companion. Before the second supernova explosion, the mass transfer that spun-up
PSR B1534+12 also almost certainly circularized the binary orbit, so we
take $e_i=0$. The orbital velocity just prior to the supernova is $\vec{V}_i$;
from Kepler's laws the (constant) orbital speed is given by $V_i^2=GM_i/a_i$.
During the explosion, mass $M_i-M_f=m_{2i}-m_{2f}$ is lost from the system, and
a velocity kick $\vec{V}_k$ (the ``natal kick'') 
is applied to the companion star. 
After the explosion, the new elements are $e_f$, $a_f$, and
$M_f=m_1+m_{2f}$, and the change in the direction of the orbital angular
momentum is $\delta$. Expressions for the postexplosion elements can be found in
\citet{hil83}. For convenience, we follow the very clear exposition of
\citet{wkk00} and define normalized quantities
$\tilde{v}=\vec{V}_k/V_i$, $\alpha\equiv a_f/a_i$, $\beta=M_f/M_i$, and
$\eta=\sqrt{\alpha\beta(1-e_f^2)}$. Then the shifts in the elements can be
related to the components of the kick velocity vector through:
\begin{mathletters}
\begin{eqnarray}
\label{eqn:vx}
\tilde{v}_x&=&\pm\eta\left[\left(1-\frac{1}{\alpha\left(1+e_f\right)}\right)
\left(\frac{1}{\alpha\left(1-e_f\right)}-1\right)\right]^{1/2}\\
\tilde{v}_y&=&\eta\cos\delta-1\\
\tilde{v}_z&=&\pm\eta\sin\delta. \label{eqn:vz}
\end{eqnarray}
\end{mathletters}
Here the $\hat x$ axis is the line connecting the pulsar and companion at the
time of the explosion, $\hat y$ is the direction of instantaneous velocity of
the companion, and $\hat z$ is perpendicular to the orbit.
From equation~\ref{eqn:vx} it is straightforward to see that the initial semimajor
axis is bounded by
\begin{equation}
\frac{1}{1+e_f}\leq\alpha\leq\frac{1}{1-e_f}.\label{eqn:alim}
\end{equation}
For any  $\alpha$ in this range, $\beta<1$, and $\delta$, we can use
equations.~\ref{eqn:vx}--\ref{eqn:vz} to calculate the kick velocity vector
needed to produce the observed system.

Given the mass loss and kick applied to the companion star, the resulting kick to the binary
center of mass $\tilde{u}=\vec{u}/V_i$ can also be calculated. We refer to this
as the ``systemic kick.'' (Note that the binary receives a systemic kick even in
a symmetric explosion, when $\vec{V}_k=0.$) 
Again, we follow \citet{kal96} and \citet{wkk00} and write
\begin{equation}
\tilde{u}^2=\kappa_1+\kappa_2\left(2-\alpha^{-1}\right)-
\kappa_3\sqrt{\alpha\left(1-e_f^2\right)}\cos\delta,\label{eqn:systemic}
\end{equation}
where
\begin{equation}
\kappa_1\equiv\frac{m_{2i}^2}{M_i^2},\quad
\kappa_2\equiv\frac{m_{2f}^2}{M_iM_f},\quad\mbox{and}\quad
\kappa_3\equiv2\sqrt{\kappa_1\kappa_2}.
\end{equation}
We can decompose this into the component perpendicular to the post-supernova
orbit $\tilde{u}_\perp$ and the component in the plane of the orbit
$\tilde{u}_\parallel$, with 
\begin{mathletters}
\begin{equation}
\tilde{u}_\perp=\sqrt{\kappa_1}\sin\delta\label{eqn:uperp}
\end{equation}
and
\begin{equation}
\tilde{u}_\parallel^2=\kappa_1\cos^2\delta+\kappa_2\left(2-\alpha^{-1}\right)-
\kappa_3\sqrt{\alpha\left(1-e_f^2\right)}\cos\delta,\label{eqn:upar}
\end{equation}
\end{mathletters}

\subsection{Evolution after the second supernova}

The current binary system consists of two neutron stars. The orbit evolves
slowly as it loses energy and angular momentum to gravitational radiation.
Neither mass transfer nor tidal effects will have any significant contribution
to the orbital evolution until the last stages of inspiral, about 3~Gyr from now.

Although high precision radio pulsar timing observations have measured
the effects of gravitational radiation damping on the orbital
elements, they are very small.  Indeed, the total change in the elements
since the second supernova explosion formed the double-neutron star binary can
for most purposes be ignored.
Following \citet{pet64}, we can write the fractional rates of change of the
semi-major axis and eccentricity as:
\begin{mathletters}
\begin{equation}
\frac{1}{a}\frac{da}{dt}=
-\frac{64G^3}{5c^5}\frac{m_1m_2M}{a^4\left(1-e^2\right)^{7/2}}
\left(1+\frac{73}{24}e^2+\frac{37}{96}e^4\right)
\end{equation}
and
\begin{equation}
\frac{1}{e}\frac{de}{dt}=
-\frac{304G^3}{15c^5}\frac{m_1m_2M}{a^4\left(1-e^2\right)^{5/2}}
\left(1+\frac{121}{304}e^2\right).
\end{equation}
\end{mathletters}

For PSR B1534+12, the timescales for evolution of $a$ and $e$ are 9.0~Gyr and
2.0~Gyr, respectively: both much longer than the pulsar characteristic age. We 
can therefore ignore post-supernova evolution of the orbital
elements. As we will see, we are actually interested only in $a$ and in $1\pm
e$. Although we can't properly correct these quantities to their original values
without knowing the true binary age, the errors that enter when we neglect
radiation damping are $\stackrel{<}{\sim}3\%$, which will be dwarfed by uncertainties in
stellar modeling.

Because tidal effects are extremely small, no gravity-driven
evolution of $\delta$ is expected. Recently, \citet{drb+04} have suggested that
wind from the fast pulsar in the PSR J0737$-$3039 binary, which penetrates
well into the light-cylinder of the slower pulsar, may have aligned the
spin axis of the slow pulsar with the orbital angular momentum. Such alignment
is unlikely for a fast pulsar like B1534+12; for any likely birth luminosity,
the standoff radius would have
been well outside the light cylinder of B1534+12. The orbital motion
dominates the total angular momentum of the system, so the orientation of the orbit
remains nearly fixed in space, while the spin axis of the pulsar precesses
around it tracing a cone with half opening angle $\delta$.

In summary, because of its small age and slow evolution, we can without
significant error take the currently observed orbital parameters, orbital orientation,
and spin-orbit misalignment to be an accurate
record of the state of the system immediately after the second supernova
explosion.
The space velocity imparted to the system during the second supernova is not,
however, preserved as the binary moves through the Galactic potential. We
discuss this complication in \S\ref{sec:kinematics}.

\section{Discussion}\label{sec:results}

Our goal is to combine the wealth of observational data on the current state of the
binary system and its position and motion through the Galaxy with constraints
that come from binary evolution modeling and the effects of the mass loss during
the supernova event to develop a consistent model of the late evolution of the
system. Several free parameters remain that must be constrained. Among the
quantities not
directly accessible from observations are the radial velocity (and hence the
full space velocity), the age and birthplace, the pre-supernova mass and radius
of the binary orbit, and the asymmetry of the supernova explosion.  
The constraints on these parameters are interrelated and
complex.

To clarify the situation, we begin by allowing the assumed radial velocity of
the system to vary. We will find that assuming the system was born near the plane of
the Galaxy (where most massive stars are found) in the last 210~Myr is
sufficient to severely limit both the current space velocity of the pulsar and
the kick velocity given to the binary system at birth. This systemic kick, the
post-supernova eccentricity and size, and the misalignment all depend on the
pre-supernova orbital size, the mass loss during the supernova, and
the magnitude and direction of any asymmetric kick given to the companion.

We will show that the range of kick magnitudes and progenitor masses that 
satisfy all the constraints is very limited.

\subsection{The binary elements}

The relatively low eccentricity of the observed binary together with equation~\ref{eqn:alim}
constrains the
size of the pre-supernova binary to within about 30\%. Evolutionary arguments
constrain the pre-supernova mass of the companion star to between about 2 and 5.5
solar masses. Precession studies constrain the post-explosion misalignment with
the discrete a ambiguity between nearly aligned pulsar spin and
orbital angular momenta ($\delta\sim25^\circ$) and nearly counteraligned angular
momenta ($\delta\sim155^\circ$). 

Given the observed orbital parameters for PSR~B1534+12, we can rewrite 
equations~\ref{eqn:vx}--\ref{eqn:vz} as 
\begin{mathletters}
\begin{eqnarray}
V_x  &=& \pm 380 
\left [(\alpha-0.785)(1.377-\alpha ) \right ]^{1/2} \\
V_y & = & 395 \left (\pm 0.872 \alpha - \sqrt{\frac{\alpha}{\beta}} \right
)\label{eqn:vyn}\\
V_z &=&  \pm 160\alpha
\end{eqnarray}
\end{mathletters}
In equation~\ref{eqn:vyn}, the $+$ sign refers to the nearly aligned case, the $-$
sign to the nearly counteraligned case.

From the observed parameters, we know that $0.785 \leq \alpha \leq 1.38$. With
no assumptions about the mass loss in the second supernova explosion, this gives
us direct constraints on two components of the kick velocity: 
$130 \leq |V_z| \leq 220$~km/s, and $|V_x| \leq 115$~km/sec.

The parameter $\beta$ is strictly less than one, reaching this limit only if no
mass is lost in the second supernova.  With the conservative assumption that
the presupernova mass of the companion
star was at least $2M_\odot$, we have $\beta\stackrel{<}{\sim}.8$. If the misalignment of the
system is $\delta\sim155^\circ$, then a very large kick in the direction
opposite to the instantaneous orbital motion is required, with $V_y<-620$~km/s
for no mass loss and $V_y<-660$~km/s for a $2M_\odot$ presupernova mass. The
minimum total kick speeds are 630~km/s and 670~km/s.

In the $\delta\sim25^\circ$ case, much smaller kicks are acceptable. For the
no-mass-loss case the constraint is $-80<V_y<11$~km/s, with a minimum total kick
speed of 149~km/s, for a $2M_\odot$ presupernova mass 
$-120<V_y<-44$~km/s with a minimum total of 173~km/s. 

\begin{figure}[t]
\plotone{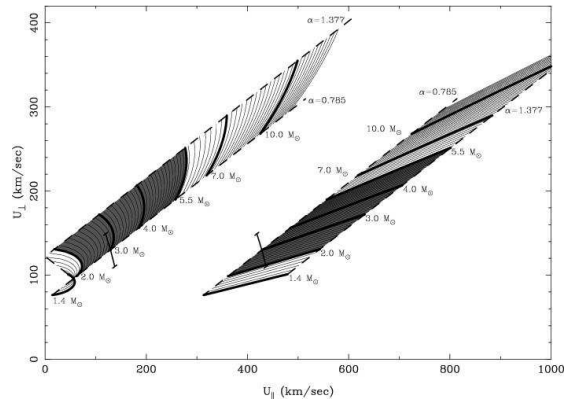}
\caption{\label{fig:butterfly}The center-of-mass (or ``systemic'') velocity
imparted to the PSR B1534+12 binary at the time of the second supernova, broken
into components perpendicular and parallel to the post-explosion orbital plane.
The sign of $U_\perp$ is unconstrained.
The shaded regions mark the initial
states allowed by the presupernova binary evolution models of \citet{dp03b}. The
``island'' of points at relatively low velocity are the solutions with
$\delta=25^\circ$, while the island at high velocity are solutions with
$\delta=155^\circ$. Representative error bars show the effect of varying
$\delta$ within the allowed $1\sigma$ range. 
Lines of constant presupernova mass are shown,
as are lines of constant presupernova orbital size, as parameterized by $\alpha$
(see eq.~\ref{eqn:alim}).
}
\end{figure}

Given the mass loss and natal kick, we can use equations~\ref{eqn:uperp} and~\ref{eqn:upar}
to calculate the systemic velocity kick acquired by the binary in the second
supernova. In Fig.~\ref{fig:butterfly}, we show the relationship between the pre-supernova
properties of the binary and the components of this kick perpendicular and
parallel to the post-explosion orbital plane. It is immediately evident that
a natal kick large enough to leave $\delta\sim155^\circ$ inevitably leaves the
binary system moving fast compared to the observed 122~km/s transverse velocity,
suggesting that such large misalignments are only possible if the current binary
velocity is fortuitously aligned very nearly along the line-of-sight to the
pulsar.  As we will see in the next section, such alignment is not consistent
with our knowledge of the orientation of the binary.

\subsection{Motion of the binary in the Galaxy}
\label{sec:kinematics}

The full three-dimensional orientation of the PSR B1534+12 orbit is known. The
orbital inclination relative to the plane of the sky ($\sin i$) is known from 
timing measurements. The usual ambiguity in the sign of $\cos i$, which is not
accessible from timing studies, is known from the profile observations. The
angle of the line of nodes, corresponding to the rotation of the binary around
the line of sight, is known from scintillation studies. Together, these
observations fix the direction of the orbital angular momentum vector in space.
Because tidal forces on the binary are very small, there has been no significant
change in the direction of the angular momentum vector since the current binary
was formed in the second supernova explosion.

The arguments of the last section give us, for any given presupernova
parameters, the components of the systemic velocity kick parallel to and
perpendicular to this angular momentum vector (e.g., Fig.~\ref{fig:butterfly}).
Ideally, we could compare those kick velocity components directly to the observed space
motion of the binary, but we must first overcome two significant problems.
First, although the transverse motion of the binary is well determined, there is 
no direct constraint on the radial velocity. Second, the current velocity has
been substantially affected by acceleration of the binary as it moves in the
Galactic gravitational potential. 

Fortunately, the age of the binary is relatively low, so it is not unreasonable
to follow its motion backwards to find possible birth locations and to thereby
estimate the systemic kick at birth.
It is very likely that the original pair of massive stars
was born near the plane of the Galaxy, moving
relatively slowly. The scale height of O type stars is small, $\sim 50$~pc
\citep{sto79}, and their peculiar velocities are low. Even after the first
supernova explosion, the space velocity remains low, comparable to the
fractional mass loss times the presupernova orbital velocity. In the canonical
evolutionary model, where conservative mass transfer prior to the first
supernova leads to orbital widening \citep[for example]{bv91}, a typical
velocity kick is $\sim 10-20$~km/s.

The velocity of the binary is therefore dominated by the recoil introduced in
the second supernova explosion, when both the fractional mass loss and the
pre-supernova orbital velocity are much higher than they were in the first explosion. 
The binary is currently far from the Galactic plane, at $z=680$pc. Assuming the binary
was moving with the local Galactic rotational velocity and was
in the Galactic plane at the time of the second supernova will
introduce only minor errors to estimates of the age and systemic kick velocity.

We have studied the motion of the pulsar through the Galaxy using the 
potential model of \citet{kg89}. A series of trial radial velocities
was used. In each case, the solar motion was added to the pulsar motion to
produce a space velocity. The orbit of the pulsar was then integrated backwards for
210~Myr. Each plane crossing represents a potential birth place and time
for the binary system. Subtraction of the local rotational velocity then yields
the systemic kick velocity needed at the time of the second supernova to carry
the binary to its observed location. This can be decomposed into components parallel
to and perpendicular to the post-supernova orbital plane, but we first
discuss conclusions that are independent of the orientation of the system.

Depending on its age, birth location, and kick velocity, we find
that the binary 
could be on its first, second, third, fourth, or even (with careful fine tuning)
even fifth
excursion away from the Galactic plane. If this is {\it not} the initial departure
from the plane, then we find that any pulsar age between 70 and 210~Myr is possible, as
is any current radial velocity between about $\pm220$~km/s.

Radial velocities with $v_r\stackrel{>}{\sim}200$~km/s are only acceptable if it is still on its initial
excursion away from the plane.  A large velocity 
directed nearly radially away from us implies that the pulsar was born recently
in the near neighborhood of the Sun. For example, $v_r=200$~km/s implies an age
of just 7.9~Myr, while $v_r=300$~km/s implies
an age of 4.5~Myr and a birth within a kiloparsec from the Sun. While we cannot rule out
arbitrarily large radial velocities, they require a fortuitous viewing angle
and we believe them to be highly unlikely for other reasons.
First, the young implied ages require that the pulsar birth period was very
close to the observed period, and well below the spin-up line. This requires
very
fine tuning of the accretion parameters during spin-up. Second, unless we were
extraordinarily lucky to find a double neutron star system born so recently in
the solar neighborhood, the implied birthrate would be extraordinarily high.

If the binary has a large radial velocity {\it
towards} us, then it must be returning to the plane for the first time. Again, fortuitous
viewing alignments are difficult to rule out. However, in this case the velocity
is directed generally towards the Galactic center, so a radial velocity as high
as $-300$~km/s implies a birthplace at a very large Galactocentric radius,
$R>15$~kpc. We regard birth at such large radii to be unlikely.

We conclude, therefore, that the radial velocity is almost certainly 
between $-300$~km/s and $+220$~km/s, and probably between $\pm220$~km/s. We
cannot set firm limits on the age of the system, but it is very likely 
more than about 100~Myr. Most important
for our present purposes, we note that in all of the acceptable models
the systemic kick given to the binary at
birth was between $\sim100$ and $\sim240$~km/s.

Note that this conclusion is already strong enough to exclude the very large
kicks needed to produce $\delta\sim155^\circ$. Also excluded are scenarios with
large presupernova masses, such as those of \citet{dp03b} that avoid a second 
period of Roche lobe overflow, since the minimum presupernova mass in these models,
$\sim4.5M_\odot$, requires a minimum systemic kick of $\sim270$~km/s.  
The only models consistent with the observed low
velocities are the case BB models in which the second helium star explodes with
a mass of $\sim2-4M_\odot$.

\subsection{Orientation of the orbit}

\begin{figure}[t]
\plotone{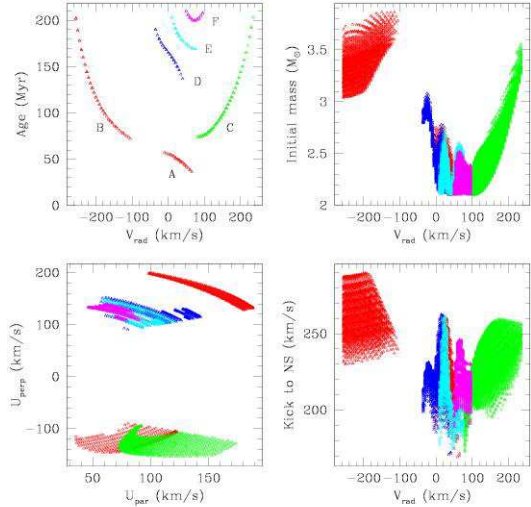}
\caption{\label{fig:allowed} Possible histories of the PSR B1534+12 binary.
For each possible radial 
velocity, the orbit in the Galaxy was integrated backwards, and each
Galactic plane crossing was recorded. The necessary systemic kick components 
parallel to and perpendicular to
the post-supernova orbit were calculated. Plotted solutions are those allowed for some choice
of presupernova companion mass $m_{2i}$ and semimajor axis $a_i$, given $1\sigma$
ranges in the parameters $\delta$ and $\Omega$. Several
families of acceptable solutions exist, marked by letters and summarized in
Table~\ref{tab:solns}. 
Top left: Possible ages of the binary system, assuming the 
binary was born near the Galactic plane.
The lowest curve in this figure (red points in the
electronic edition) includes possible
histories where the binary is leaving the plane for the first time (positive radial
velocities) or returning for the first time (negative radial velocities).
The second curve (green in the electronic edition) describes solutions where 
the binary is leaving or returning to the
plane for the second time, and so forth. For the smallest allowed orbits, there is
time in 210~Myr for the pulsar to be on its fifth excursion from the plane (having
started its third complete orbit in the Galactic potential). 
Top right: Allowed presupernova companion masses. 
Bottom left: Allowed systemic kick velocities, relative to the postsupernova orbital
plane. This figure can be compared with Fig.~\ref{fig:butterfly}. Note that there are
no solutions consistent with $\delta\approx155^\circ$.
Bottom right: Magnitude of the asymmetric velocity kick imparted to the companion star at
the time of the second supernova explosion. }
\end{figure}

\begin{figure}[t]
\plotone{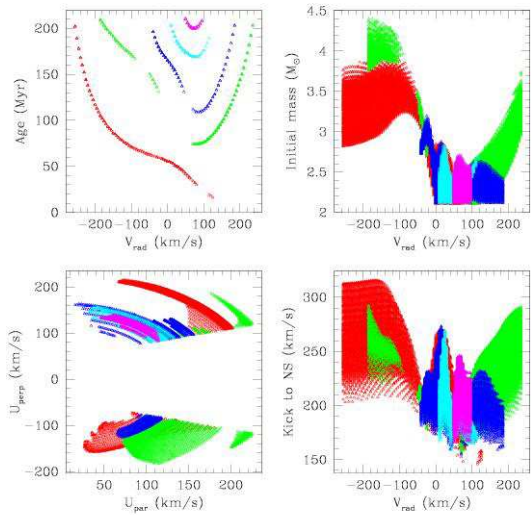}
\caption{\label{fig:allowedb} Identical to Fig.~\ref{fig:allowed}, except
$2\sigma$ ranges in $\delta$ and $\Omega$ are accepted. The most important
difference is that some second-pass solutions with larger negative radial
velocity are allowed, somewhat broadening the range of allowed presupernova
companion masses.}
\end{figure}

\begin{deluxetable}{ccrcc}[h]
\tablecolumns{5}
\tablewidth{0pt}
\tablecaption{Possible birth properties\label{tab:solns}}
\tablehead{
\colhead{Key\tablenotemark{a}} 
& \colhead{Pass\tablenotemark{b}}
& \colhead{Age\tablenotemark{c}} 
& \colhead{$m_{2i}$} 
& \colhead{$V_k$} \\
&&\colhead{(Myr)} & \colhead{($M_\odot$)} &  \colhead{(km/s)}
}
\startdata
A & 1 & 40--60   & 2.1--2.7 &  170--260 \\
B & 1 & 70--210  & 3.0--3.9 &  230--290 \\
C & 2 & 70--210  & 2.1--3.6 &  170--260 \\
D & 3 & 130--190 & 2.1--3.1 &  170--260 \\
E & 4 & 170--210 & 2.1--2.7 &  170--260 \\
F & 5 & 200--210 & 2.1--2.5 &  170--250 \\
\enddata
\tablenotetext{a}{See Fig.~\ref{fig:allowed}.}
\tablenotetext{b}{Excursion from the Galactic plane. For example, pass~2 means
that the binary has already returned to the plane once since receiving its large
velocity kick during the second supernova.}
\tablenotetext{c}{Time since second supernova explosion.}
\end{deluxetable}

These constraints can be further tightened  using our knowledge of the
post-supernova orbital orientation. After propagating the binary back to a
potential birthplace to find the direction and magnitude of the systemic kick
that was needed at birth, we can resolve this kick into components perpendicular
and parallel to the post-supernova orbital plane and solve equations~\ref{eqn:systemic}
and~\ref{eqn:uperp} for the presupernova orbital size and companion mass. Using
Kepler's law to eliminate $a_i$, and some algebraic manipulation, 
we can rewrite equation~\ref{eqn:systemic} as a simple
quadratic equation in $m_{2i}$: 
\begin{eqnarray} 
0 &=&  m_{2i}^2
\left[u^2-\frac{u_{\perp}^2}{\sin^2\delta} +\frac{Gm_{2f}^2}{(m_{2f}+m_1)a_f}
\right]\nonumber\\
 & &-{}2(m_{2i}+m_1) \left [
\frac{m_{2f}^2}{m_{2f}+m_1}\frac{u_{\perp}^2}{\sin^2\delta} \right. \nonumber\\
 & & \left. -{}m_{2f}\frac{u_{\perp}^3}{\sin^3\delta}\sqrt{\frac{a_f}{G(m_{2f}+m_1)}}\sqrt{1-e^2}
\cos\delta \right ] 
\end{eqnarray} 
Allowing for the sign ambiguity in $u_\perp$,
there are up to four real roots, each with a
corresponding $a_i$ that can be determined from Kepler's equation.

Physically interesting solutions have $m_{2i}$ in the range
$2.1\,M_{\odot}$--$10\,M_{\odot}$ and $a_i$ within the range given by
equation~\ref{eqn:alim}. We consider 1~and $2\sigma$ ranges in $\Omega$ and $\delta$ when
computing the full set of possible progenitors. The resulting set of allowed
solutions are shown in Figs.~\ref{fig:allowed} and~\ref{fig:allowedb}. 
We find viable solutions with
1--5 disk crossings and ages less than 210 Myr. There are no solutions
with radial velocities less than $-260$\,km/s or more than 240\,km/s. 
The kick to the NS is relatively tightly constrained to between
170 and 290\,km/s (between 150~km/s and 320 allowing for $2\sigma$ regions), 
while the progenitor mass cannot have been more than
$\sim3.8\,M_{\odot}$ ($4.4M_\odot$ for $2\sigma$ regions), 
again implying that only case BB progenitors are possible.
The kick direction is oriented within roughly $20$--$40^{\circ}$ or
$140$--$160^{\circ}$ of the pre-supernova orbital angular momentum (with the
ambiguity again arising from the ambiguity in the sign of the perpendicular
component of the velocity kick) and the component of the natal kick that is in
the presupernova orbital plane is directed roughly opposite to the instantaneous
progenitor motion at the time of explosion.

Several sources of uncertainty contribute to our results at a comparable level.
Included in our analysis are the $\sim20^\circ$ uncertainty in $\Omega$, which
introduces uncertainty in the
decomposition of the birth velocity into parallel and perpendicular components,
and the $\sim4^\circ$ uncertainty in $\delta$ (e.g., Fig.~\ref{fig:butterfly}). A
smaller contribution is the $10-20$~km/s uncertainty in
the presupernova peculiar velocity of the binary.

Another significant uncertainty, particularly for solutions with high space velocity,
is the Galactic potential model used to model the binary's motion from the
Galactic plane. We have repeated our analysis of the binary space motion
using the three component potential of \citet{jsh95}, and find very similar results.
The only significant difference is that for a very small range of radial velocities
(between $-205$ and $-220$~km/s, on the second excursion from the Galactic plane,
ages 200--210~Myr)
there are models with more massive presupernova progenitors ($m_{2i}=5$--$5.6M_\odot$)
that are consistent with the errors in $\Omega$ and $\delta$. We regard these
scenarios as unlikely, requiring careful fine-tuning as well as less-commonly
used potential model, but the difference does highlight the systematic
uncertainties in following the orbit of such high-velocity objects in the
Galaxy. (Those particular solutions are just returning to the plane from a
maximum height of $\sim6$~kpc.)

In summary, a number of error sources as well as discrete ambiguities in the age
of the system make it impossible to precisely determine the parameters of the
presupernova orbit, even with the available data on the orbital orientation. 
However, the data are sufficient to strongly constrain
the parameters: it is, for example, likely that the progenitor 
mass was in the range
$\sim3\pm1M_\odot$, and that the natal kick is $\sim230\pm60$~km/s.

\section{Conclusions}
The combination of timing, precession, and proper motion studies have allowed us
to strongly constrain the evolution of the PSR B1534+12 binary. There are
several conclusions that can be drawn:
\begin{enumerate}
\item The misalignment of the pulsar spin axis from the orbital angular momentum
is relatively small, with $\delta=25.0\pm3.8^\circ$. Although a 
misalignment of $155.0\pm3.8^\circ$ was formally allowed by observations of
geodetic precession in the system, such a large misalignment would require an
asymmetric kick at the time of the second supernova explosion that is too large
to reconcile with the current space velocity of the binary.
\item Of the evolutionary scenarios studied by \citet{dp03b}, only one is
consistent with the observed properties and space velocity of the binary. 
The pulsar was
the first-born neutron star of the pair. After an initial inspiral during either
case B or C mass transfer as its companion evolved off the main sequence, 
the companion was a helium star between about 3.3 and $6.5M_\odot$. A
second phase of mass transfer (case BB) left the companion with mass
$3\pm1M_\odot$ at the time of the second supernova explosion. Models that avoid 
a second phase of Roche-lobe-overflow-driven
mass transfer in
the binary by positing a high zero age helium main sequence mass for the
companion also require an asymmetric kick that is too high.
\item
In the second supernova explosion, an asymmetric kick of
$230\pm60$~km/s was required to produce the observed orbital elements. The
direction of the kick relative to the orbital angular momentum vector (and
presumably the angular momentum of the presupernova core) is tightly
constrained, and is not obviously biased towards either the pole or the orbital
plane.
We note that the inferred asymmetric kick is somewhat larger than the 100-150~km/s 
most-likely kick for PSR J0737$-$3039,
estimated from scintillation \citep{rkr+04} and evolutionary arguments
\citep{wk04}, and probably somewhat smaller than the kick given to the companion
of PSR B1913+16, which is poorly constrained. An early analysis found 
most likely solutions for PSR B1913+16 to be $\sim300-500$~km/s
\citep[][prograde solution]{wkk00}, but the range expands to 190--600~km/s when
evolutionary scenarios including Roche lobe overflow are included \citep{wkh04}. 
The limited measurement
quality and complex selection biases make it impossible to estimate the
distribution of kick velocities imparted to newborn neutron stars in close
binaries, but we note that there is no significant evidence in this small sample
for a different velocity distribution.
\end{enumerate}

\acknowledgements
I.H.S. holds an NSERC University Faculty Award and is further supported by a
Discovery Grant.
R.J.D. and S.E.T. are supported by
the NSF under grant AST-0098343. We thank Zaven Arzoumanian and Joseph Taylor
for significant contributions to the observations on which this work is based,
and Vicky Kalogera for interesting discussions. 


\end{document}